# The Subatomic Particle Mass Spectrum

Robert L. Oldershaw

12 Emily Lane

Amherst, MA 01002

USA

rloldershaw@amherst.edu

**Abstract:** Representative members of the subatomic particle mass spectrum in the 100 MeV to 7,000 MeV range are retrodicted to a first approximation using the Kerr solution of General Relativity. The particle masses appear to form a restricted set of quantized values of a Kerr-based angular momentum-mass relation: $M = n^{1/2} \mathfrak{M}$, where values of n are a set of discrete integers and $\mathfrak{M}$ is a *revised* Planck mass. A fractal paradigm manifesting global discrete self-similarity is critical to a proper determination of $\mathfrak{M}$, which differs from the conventional Planck mass by a factor of roughly $10^{19}$. This exceedingly simple and generic mass equation retrodicts the masses of a representative set of 27 well-known particles with an average relative error of 1.6%. A more rigorous mass formula, which includes the total spin angular momentum rule of Quantum Mechanics, the canonical spin values of the particles, and the dimensionless rotational parameter of the Kerr angular momentum-mass relation, is able to retrodict the masses of the 8 dominant baryons in the 900 MeV to 1700 MeV range at the < 99.7% > level.

"There remains one especially unsatisfactory feature [of the Standard Model of particle physics]: the observed masses of the particles, m. There is no theory that adequately explains these numbers. We use the numbers in all our theories, but we do not understand them – what they are, or where they come from. I believe that from a fundamental point of view, this is a very interesting and serious problem."                                           Richard Feynman

## 1. Introduction

Retrodictions and predictions of subatomic particle masses have been highly valued *desiderata* ever since these unanticipated ultra-compact systems were discovered empirically. It is widely acknowledged that critical particle masses have to be put into the Standard Model of particle physics "by hand", and further, that this lack of predictive/retrodictive capability is considered to be a significant problem that eventually must be resolved.

In this paper we consider an unorthodox but promising approach to addressing the enigma posed by the particle mass spectrum. The main underlying idea of the approach is that gravitational interactions are much stronger *within* subatomic particles than was previously realized, by a factor of ~ $10^{38}$. We explore the hypothesis that these ultra-compact subatomic particles can be approximated as quantized allowed values, i.e., eigenstates or excited states, of the basic angular momentum-mass relation of the Kerr solution of the Einstein field equations of General Relativity. In Section 3 the justification for this "strong gravity" approach will be discussed.

## 2. The Kerr Solution In The Subatomic Domain

For the Kerr solution of the Einstein field equations of General Relativity one finds the following simplifying relationship[1] between the angular momentum (J) of an ultra-compact object and its mass (M):

$$J = \mathbf{a}GM^2/c \ . \qquad (1)$$

The parameter **a** is referred to as a dimensionless spin parameter associated with the rotational properties of the ultra-compact object, G is the gravitational coupling factor and c is the velocity of light. Since we are interested in the masses of the ultra-compact objects, we rewrite Eq. (1) in the form:

$$M = (Jc/\mathbf{a}G)^{1/2} \ . \qquad (2)$$

Since we are interested in applying Eq. (2) in the subatomic domain, we hypothesize that the unit of J in this domain is ℏ, and in the subatomic realm we expect J to be restricted to a discrete set of values, i.e., nℏ. As is well-known, in the 1960s Tullio Regge demonstrated that the masses and total spins of families of baryon and meson resonances were related by $J = kM^2$ relations.[2-4] While Regge's heuristic phenomenology is well-documented, it has not be given a fully adequate explanation within QCD or any other part of the Standard Model of particle physics.

With the above assumptions concerning J, and making the preliminary assumption that **a** = 1.00 which designates the maximum allowed rotation, we can rearrange Eq. (2) to yield:

$$M = n^{1/2} (\hbar c/G)^{1/2} \ . \qquad (3)$$

We notice that $(\hbar c/G)^{1/2}$ is just the definition for the Planck mass. Therefore according to Eq. (3) the allowed values of the Kerr-derived J versus $M^2$ relation in the subatomic domain are the square roots of quantized multiples of the Planck mass. The applicability of the Kerr solution of General Relativity in the subatomic realm, and our initial assumptions concerning **a** and J, can be tested by attempting to retrodict the subatomic particle mass spectrum using Eq. (3).

## 3. Evaluating $(\hbar c/G)^{1/2}$ With Discrete Scale Relativity

The first step in testing Eq. (3) in the subatomic domain is to re-evaluate the Planck mass. Motivation for questioning the conventional Planck mass can be found in: (1) the fact that the conventional value of 2.176 x $10^{-5}$ g is not associated with any particle or phenomenon observed in nature, (2) the fact that this mass value results in many forms of the closely related "hierarchy problem", and (3) the fact that it leads to a vacuum energy density (VED) crisis in which there is a disparity of 120 orders of magnitude between the VED estimates of particle physics and cosmology.

A way to avoid these problems, and quite a few more, can be found in a new cosmological paradigm for understanding nature's structural organization and dynamics. This new fractal paradigm is called the Discrete Self-Similar Cosmological Paradigm (DSSCP). It is the product of a very thorough and careful empirical study of the actual objects that comprise nature, and the paradigm is based on the fundamental principle of discrete scale invariance. The discrete self-similar systems that comprise nature, and the fact that fractal structures are so common in nature, are the physical manifestations of the discrete scale invariance of nature's most fundamental laws and structural geometry. The DSSCP was reviewed in two published papers[5,6] and a comprehensive website is available at http://www3.amherst.edu/~rloldershaw.

Discrete Scale Relativity (DSR) is the variation of the general DSSCP which postulates that nature's discrete self-similarity is *exact*, as discussed recently in a brief paper.[7] According to DSR, the gravitational coupling factor scales in the following discrete self-similar manner:

$$G_\Psi = (\Lambda^{1-D})^\Psi G_0 ,  \qquad (4)$$

where $G_0$ is the conventional Newtonian gravitational constant, $\Lambda$ and $D$ are empirically determined dimensionless self-similarity constants equaling $5.2 \times 10^{17}$ and $3.174$, respectively, and $\Psi$ is a discrete index denoting the specific cosmological Scale under consideration. For the evaluation of Eq. (3) we have $\Psi = -1$, which designates the Atomic Scale, and therefore $G_{-1} = \Lambda^{2.174} G_0 = 2.18 \times 10^{31}$ cm$^3$/g sec$^2$. According to DSR, $G_{-1}$ is the proper gravitational coupling constant between matter and space-time geometry *within* Atomic Scale systems. Evaluating the Planck mass relation $(\hbar c/G)^{1/2}$ using $G_{-1}$ and the usual values of $\hbar$ and c yields a value of $1.203 \times 10^{-24}$ g, corresponding to 674.8 MeV. This revised Planck mass is identified below by the symbol ℳ. To a first approximation, the subatomic particle mass spectrum should have peaks at the mass-energy values:

$$M = n^{1/2} \mathcal{M} = n^{1/2} (674.8 \text{ MeV}) . \qquad (5)$$

## 4. Testing M = n$^{1/2}$ 𝔐

Table 1 presents relevant data for testing Eq. (5) in terms of a *representative* set of subatomic particles from a mass-energy range of 100 MeV to 7,000 MeV. The particles appearing in Table 1 are among the most abundant, well-known and most stable members of the particle/resonance "zoo". For each integer of n there appears to be an associated particle, or set of related particles, that agrees with the quantized mass values generated by Eq. (5) at about the 93 to 99.99 % level. The *average* relative error for the full set of 27 particles is 1.6 %.

**Table 1**

**Representative Subatomic Particle Mass Spectrum (100 MeV to 7,000 MeV)**

| n | n$^{1/2}$ | n$^{1/2}$ (674.8 MeV) | Particle / MeV | Relative Error |
|---|---|---|---|---|
| 1/36 = (1/9)/4 | 0.1666 | 112.46 | μ / 105.66 | 6.4 % |
| 1/25 ≈ (1/6)/4 | 0.2000 | 134.96 | π / 134.98 | 0.01 % |
| 1/2 = 2/4 | 0.7071 | 477.15 | κ / 497.65 | 4.1 % |
| 3/4 | 0.8660 | 584.39 | η / 547.75 | 6.7 % |
| 1 = 4/4 | 1.0000 | 674.8 | 𝔐 / 674.8 | --- |
| 5/4 | 1.1284 | 761.40 | ρ / ~ 770 | 1.1 % |
| 5/4 | 1.1284 | 761.40 | ω / ~ 783 | 2.8 % |
| 2 | 1.4142 | 954.31 | p$^+$ / 938.27 | 1.7 % |
| 2 | 1.4142 | 954.31 | n / 939.57 | 1.6 % |
| 2 | 1.4142 | 954.31 | η' / 957.75 | 0.4 % |
| 3 | 1.7320 | 1167.75 | $\Lambda^0$ / 1115.68 | 4.7 % |
| 3 | 1.7320 | 1167.75 | $\Sigma^i$ / <1192> | 2.0 % |
| 4 | 2.0000 | 1349.60 | $\Xi^0$ / 1314.83 | 2.6 % |
| 5 | 2.236 | 1508.90 | N(1440)/ 1430-1470 | ~ 4.8 % |
| 6 | 2.4495 | 1652.91 | $\Omega^-$ / 1672.45 | 1.2 % |
| 7 | 2.6458 | 1785.35 | $\tau^-$ / 1784.1 | 0.05 % |
| 8 | 2.8284 | 1908.62 | $D^0$ / 1864.5 | 2.4 % |
| 8 | 2.8284 | 1908.62 | $D^{+/-}$ / 1869.3 | 2.1 % |
| 8 | 2.8284 | 1908.62 | $^2H$ / 1889.77 | 1.0 % |
| 10 | 3.1623 | 2133.90 | $D_s^i$ / 2112.1 | 1.0 % |
| 12 | 3.4641 | 2337.58 | $\Lambda_c^i$ / 2284.9 | 2.3 % |
| 14 | 3.7417 | 2524.87 | $\Xi_c^i$ / <2522.75> | ~ 0.1 % |
| 16 | 4.0000 | 2699.20 | $\Omega_c^0$ / 2697.5 | 0.1 % |
| 18 | 4.2426 | 2862.93 | $^3H$ / 2829.87 | 1.2 % |
| 18 | 4.2426 | 2862.93 | $^3He$ / 2829.84 | 1.2 % |
| 30 | 5.4772 | 3696.03 | $^4He$ / 3727.38 | 0.9 % |
| 64 | 8.000 | 5398.40 | $B_j^i$ / <5313.25> | ~ 1.6 % |
| 90 | 9.4868 | 6401.71 | $B_c^i$ / <6400> | ~ 0.1 % |

Table 1 lists the n values, the retrodicted masses, the empirical masses, and the relative errors for 27 subatomic particles. Here we will discuss these 27 test particles in two separate groups: those particles that have masses $\geq m_p$, where $m_p$ is the proton mass, and those particles that have masses $< m_p$. For the former group we see that integer n-values generate good first approximation retrodictions with < 98.4 % > accuracy for the particles with masses in the $m_p \leq M < 7,000$ MeV range.

For the much smaller group of particles with $M < m_p$, the set of n-values is not as simple and regular as it is for the $M \geq m_p$ group. The unit ℳ obviously has n = 1 but other members of this group have fractional values of n. The μ, π, κ, η, ℳ, ρ and ω particles can be *assigned* n = (1/9)/**4**, (1/6)/**4**, 2/**4**, 3/**4**, 4/**4**, 5/**4** and 5/**4**, respectively, or n = 1/36, 1/25, 1/2, 3/4, 1, 5/4 and 5/4. One gets the definite impression that there is an underlying order to this subset of n-values, but a unique pattern is not obvious. The distinct possibility exists that n-values for the $M < m_p$ group are *compound terms* such as n = i / j, or n = i · j, where i and j could be integers, and/or multiple rational fractions, e.g., $n_\eta = [3/2 \cdot 1/2]$. Rather than explore these possibilities numerologically, an approach with a long and checkered history, it seems more prudent to wait for a second approximation analysis of the subatomic particle mass spectrum using the full *Kerr-Newman* solution of the Einstein-Maxwell equations to provide a more sophisticated model of the particles. This more complete and rigorous analysis would include *charge*, mass, electrodynamic considerations and spin-related phenomena. The results of this second approximation analysis should provide considerable guidance in understanding the most appropriate set(s) of n-values for all particles, as well as fostering an understanding the more subtle properties of the underlying order that generates the very regular patterns observed in the particle mass spectrum.

## 5. A More Rigorous Mass Formula

One preliminary way to refine our mass formula and make it less heuristic is to relax the restriction to *extremal* Kerr ultracompacts with **a** = 1.00. In the general case, **a** can vary between 0.00 (no rotation) and 1.00 which designates the maximal rotation for a *stable* Kerr ultracompact. Since we are modeling quantum particles, we will allow **a** values to vary between 0.00 and 1.00, with the values having the form x/y where x and y are integers and y > x.

Another important refinement to our first approximation modeling of subatomic particles using the Kerr metric will be to adopt a more rigorous expression for the total spin angular momentum of a particle:

$$J = (j\{j+1\})^{1/2}\, \hbar \ . \qquad (6)$$

Using this formal definition of J from Quantum Mechanics in place of n$\hbar$ in Eq. (3), our refined mass formula becomes:

$$M = (j\{j+1\}/\mathbf{a}^2)^{1/4}\, (674.8 \text{ MeV}) \ . \qquad (7)$$

We will only allow the conventionally assigned *canonical* j values for specific particles, and **a** values will be restricted to rational fractions that are consistent with Kerr-based stability limits.

Using Eq. (7) as our more sophisticated mass formula rules out retrodicting the masses of particles with j = 0, such as spin 0 mesons. In fact, it is very reasonable to expect that the Kerr metric approach to modeling subatomic particles is most effective in retrodicting the masses of baryons, which have j > 0 and are not "point-like" particles as is the case with leptons.  Table 2 lists the data for 8 of the most well-known and relatively stable baryons.

**TABLE 2    DATA FOR THE PRIMARY BARYONS (900 MeV to 1700 MeV)**

| Particle(s) | j | a | Retrodicted Mass (MeV) | Empirical Mass (MeV) | Relative Error |
|---|---|---|---|---|---|
| Proton (+) | 1/2 | 4/9   (~1/2) | 941.96 | 938.3 | 0.4 % |
| Neutron (0) | 1/2 | 4/9   (~1/2) | 941.96 | 939.6 | 0.3 % |
| Lambda (0) | 1/2 | 6/19   (~1/3) | 1117.48 | 1115.7 | 0.2 % |
| Sigma (+,-,0) | 1/2 | 5/18   (~1/3) | 1191.49 | < 1193.1 > | < 0.1 % > |
| Delta (++,+,0,-) | 3/2 | 7/12   (~1/2) | 1229.49 | < 1232.0 > | < 0.2 % > |
| Xi (0,-) | 1/2 | 2/9   (~1/5) | 1332.13 | < 1318.3 > | < 1.0 % > |
| Xi (0,-;1530) | 3/2 | 3/8 | 1533.44 | < 1533.4 > | < 0.003 % > |
| Omega (-) | 3/2 | 5/16   (~1/3) | 1679.80 | 1672.5 | 0.4 % |

With Eq. (7) we are able to retrodict the masses of this archetypal set of baryons at the < 99.67 %

> level.  Technically there are 15 distinct particles in this set of baryons, but it appears that the

Sigma, Delta and Xi subsets are "fine structure" variations on a "base" particle, given the closeness of the masses in each subset.  It will be interesting to explore the hypothesis that the full Kerr-Newman metric solutions will offer unique explanations for this type of fine structure in the mass-stability spectra.

## 6. Implications

The results presented in Table 1 suggest that the subatomic particle mass spectrum manifests a simple, consistent and orderly pattern extending over a considerable range of particle masses and a diversity of family types, i.e., leptons, mesons, and baryons. To a first approximation the masses and angular momenta of the particles appear to be the primary or dominant physical determinants of the particle mass spectrum. Charge and other physical phenomena appear to be second order effects that determine the fine structure of the mass spectrum.

A critical factor in retrodicting the approximate particle masses is the revised Planck mass ($\mathfrak{M} \approx 674.8$ MeV) which is uniquely obtained via the scaling relations of the Discrete Self-Similar Cosmological Paradigm. With the simple heuristic mass formula: $M = (n)^{1/2} \mathfrak{M}$, we can retrodict the masses of 27 representative particles in the 100 MeV to 7,000 MeV range at the < 98.4 % > level. With the more rigorous mass formula: $M = (j\{j+1\}/\mathbf{a}^2)^{1/4} \mathfrak{M}$, we can successfully retrodict the masses of 8 major baryons in the 900 MeV to 1700 Mev range at the < 99.7 % > level, and the analysis has been constrained by using canonical j values and reasonable values of **a** that obey the Kerr solution restrictions.

A more exact retrodiction of the particle mass spectrum will clearly require a full Kerr-Newman solution of the Einstein-Maxwell equations in order to include charge and charge-related phenomena in the analyses. It can be *definitively predicted* on the basis of the results discussed in this paper that the full Kerr-Newman solution will permit a much more accurate retrodiction of the mass spectrum that includes more of the spectrum's fine structure, such as the small mass difference between the proton and the neutron, or the slightly different masses of the $\Sigma^+$, $\Sigma^-$ and $\Sigma^0$ particles. For spin = 0 particles like pions and kaons, the Reissner-Nordstrom metric would seem to be the most reasonable modeling choice. The geometrodynamics approach[8] to working with the Kerr-Newman solution, as developed by Misner, Thorne and Wheeler, would seem to offer a simple method for conducting initial tests of these predictions. Interested readers are strongly encouraged to participate in this effort.

The significant agreement between our theoretical retrodictions and the empirical mass data argues that "strong gravity" and Discrete Scale Relativity offer a radical new way of understanding subatomic particles and particle mass-stability spectra.

# References

1. McClintock, J.E., Shafee, R., Narayan, R., Remillard, R.A., Davis, S.W. and Li, L.-X., "The Spin of the Near-Extreme Kerr Black Hole GRS 1915+105," *Astrophys. J.* **652**, 518-539 (2006).

2. Regge, T., "Introduction to complex orbital momenta," *Nuovo Cim.* **14**, 951-976 (1959).

3. Eden, R.J., "Regge poles and elementary particles," *Rep. Prog. Phys.* **34**, 995-1053 (1971).

4. Irving, A.C. and Worden, R.P., "Regge Phenomenology," *Phys. Repts.* **34**, 117-231 (1977).

5. Oldershaw, R.L., "The Self-Similar Cosmological Model: Introduction And Empirical Tests," *Internat. J. Theor. Phys.* **28**, 669-694 (1989).

6. Oldershaw, R.L., "The Self-Similar Cosmological Model: Technical Details, Predictions, Unresolved Issues, And Implications," *Internat. J. Theor. Phys.* **28**, 1503-1532 (1989).

7. Oldershaw, R.L., "Discrete Scale Relativity," *Astrophys. Space Sci.* **311**, 431-433 (2007); [DOI: 10.107/s10509-007-9557-x]; also available at http://arxiv.org as arXiv:physics/0701132v3.

8. Misner, C.W., Thorne, K.S. and Wheeler, J.A., *Gravitation*, W.H. Freeman, San Francisco, 1973.

**Acknowledgement:** I would like to thank Dr. Jonathan Thornburg for helpful suggestions regarding the technical presentation of this research.